\documentstyle[12pt]{article}
\input psfig

\newcommand{\be}{\begin{equation}}
\newcommand{\ee}{\end{equation}}
\newcommand{\bea}{\begin{eqnarray}}
\newcommand{\eea}{\end{eqnarray}}
\newcommand{\half}{{\scriptstyle{{1\over 2}}}}
\newcommand{\quart}{{\scriptstyle{{1\over 4}}}}
\newcommand{\Tr}{\mbox{\,Tr}}
\newcommand{\tr}{\mbox{\,tr}}
\newcommand{\mod}{\mbox{\,mod\,}}
\newcommand{\RE}{{\Re e}}

\begin{document}
\hfill INLO-PUB-4/98
\vskip1cm
\begin{center}
{\LARGE{\bf{\underline{Exact T-duality between Calorons}}}}\\
{\LARGE{\bf{\underline{and Taub-NUT spaces}}}}\\
\vspace{7mm}
{\large Thomas C. Kraan\footnote{e-mail: 
(tckraan,vanbaal)@lorentz.leidenuniv.nl}
and Pierre van Baal
} \\
\vspace{5mm}
Instituut-Lorentz for Theoretical Physics, University of Leiden,\\
PO Box 9506, NL-2300 RA Leiden, The Netherlands.
\end{center}
\vspace*{5mm}{\narrower\narrower{\noindent
\underline{Abstract:} 
We determine all $SU(2)$ caloron solutions with topological charge one and
arbitrary Polyakov loop at spatial infinity (with trace $2\cos(2\pi\omega)$), 
using the Nahm duality transformation and ADHM. By explicit computations we 
show that the moduli space is given by a product of the base manifold 
$R^3\times S^1$ and a Taub-NUT space with mass $M=1/\sqrt{8\omega(1-2\omega)}$,
for $\omega\in[0,\half]$, in units where $S^1=R/Z$. Implications for 
finite temperature field theory and string duality between Kaluza-Klein and 
H-monopoles are briefly discussed.}\par}
\section{Introduction}

Properties of self-dual solutions to the Yang-Mills equations of motion
have played an important role in understanding both the physical and
mathematical properties of gauge theories. The last couple of years these
solutions also feature prominently in the description of dualities in 
supersymmetric theories and in string theories, in particular for extensions 
to D-branes and M-theory.

We will present the calorons~\cite{HaSh}, which are instantons at finite 
temperature defined on $R^3\times S^1$, in an explicit and simple form for 
topological charge one. The reader interested in the physical applications, 
like for finite temperature field theory, should skip the mathematically 
oriented introduction below and go directly to section 2. Sections 3 and 4 can 
be skipped as well. A more detailed description will be published elsewhere.

We were inspired to pursue the case of calorons by a question posed one 
year ago by J. Gauntlett concerning the moduli space of calorons with 
non-trivial asymptotic behaviour of the Polyakov loop (non-trivial 
holonomy)~\cite{GaHa}. These calorons appear as the non-trivial component 
of H-monopoles~\cite{RoWi}. Oddly enough explicit solutions with non-trivial 
holonomy were not known. With trivial holonomy they can be obtained as an 
infinite periodic array of instantons, all oriented parallel in group 
space~\cite{HaSh}. Recent work on the T-duality between Kaluza-Klein and 
H-monopoles in string theory~\cite{GrHaMo} made us aware that our construction 
explicitly provides the classical duality transformation. It can be formulated 
without their embedding in string theory. We nevertheless hope this result can 
contribute to resolving some of the puzzles that seem to be involved in the 
relevant string dualities. There will be many experts better equipped than we 
are in addressing these stringy issues. 

The Nahm transformation~\cite{Nahm,NaCal}, also known as Mukai 
transformation~\cite{Mukai} when considered as a mapping between holomorphic 
vector bundles, maps self-dual fields on $R^4/\Lambda$ to self-dual fields on 
$R^4/\Lambda^*$. Here $\Lambda$ is an integer lattice and $\Lambda^*$ 
is its dual. For the gauge group $U(N)$ this Nahm transformation interchanges 
the rank $N$ and the topological charge (also mapping the first Chern class to 
its Hodge dual), as follows from a family index theorem~\cite{BrBa}. 
The family parameter is defined in terms of the moduli space of flat $U(1)$ 
connections, $\omega_z=2\pi iz_\mu dx_\mu$, which when added to the self-dual 
$U(N)$ connection does not change the curvature. This gives rise to a family 
of zero-modes for the chiral Dirac operator (the Weyl operator). The vector 
bundle defined over the (dual) space of flat connections thus obtained, has 
itself a self-dual connection. Monopoles, calorons, and instantons on $R^4$ 
can all be considered to arise from suitably chosen limits of lattices 
$\Lambda$. In particular for $R^4$ seen as a torus, all of whose sides are 
sent to infinity, the dual space is a single point (periods $L$ are mapped to 
$1/L$ and it is in this sense that the Nahm transformation is in fact a 
T-duality mapping~\cite{Tdual}). This explains the algebraic nature of the 
Atiyah-Drinfeld-Hitchin-Manin (ADHM) construction~\cite{ADHM} and can be used 
as most elegant and straightforward derivation~\cite{NaCal,CoGo,DoKr}. It can 
be shown that the Nahm transformation is an involution; applied twice it gives 
the identity operation. Furthermore it preserves the metric and hyperK\"ahler 
structure of the moduli spaces~\cite{BrBa}.

In the process of sending certain periods to infinity, boundary terms arise 
that destroy the self-duality of the Nahm bundle, but this can be repaired by 
suitably extending the Weyl operator on the dual space~\cite{NaCal,CoGo,Buck}. 
A particularly interesting feature arises on non-compact four dimensional 
manifolds for which infinity has the topology of $T^d$, where $d$ can be 
either 1, 2, or 3. These correspond respectively to instantons on 
$R^3\times S^1$, $R^2\times T^2$ and $T^3\times R$. Note that in the latter 
case infinity actually contains two disconnected three dimensional 
tori~\cite{Buck}. As the solutions have finite action the connection at 
infinity is flat, parametrised by the Polyakov loops winding around the $d$ 
circles. Combined with the flat $U(1)$ connection that is added to perform the 
Nahm transformation, the Weyl operator reduced to the asymptotic $T^d$ 
generically has a gap. This is guaranteed to be the case as long as the 
combined flat $U(N)$ connection on $T^d$ is without flat factors~\cite{DoKr}, 
i.e. does not reduce to $U(1)\oplus U(N-1)$ with $U(1)$ trivial. For $SU(2)$ 
it is easily seen that the Weyl operator reduced to the asymptotic $T^d$ will 
have a zero eigenvalue at $2^d$ values of $z$. If the holonomy in the direction
$i$ is in the center of the gauge group, the values for the components $z_i$ of 
these points will coincide. As soon as (at least one of) the Polyakov loops is 
non-trivial, the symmetry is spontaneously broken to $U(1)$. The zero-modes of 
the {\em reduced} Weyl operator lead to non-exponential decay of the 
zero-modes for the {\em full} Weyl operator and a partial integration, 
required in computing the curvature of the Nahm bundle, will pick up a 
boundary term. These boundary terms can only occur at the $2^d$ points 
mentioned above, and therefore are distributions, indeed for the calorons 
easily seen to be delta functions~\cite{Buck}. This was already realised long 
ago by Nahm himself~\cite{NaCal}, but up to now this has not led to explicit 
construction of solutions. 

While finishing this paper we became aware of ref.~\cite{Lee} in which 
some of the same issues are addressed.

\section{The solutions}

For calorons we compactify the time direction by periodic identification.
One requires the gauge fields to be periodic up to a gauge transformation.
By a suitable choice of gauge, where $A_0$ tends to zero at infinity, and
the topological charge $k$ is realised by the winding number of the gauge 
transformation that describes $A_i$ at spatial infinity, one has 
\be
A_\mu(x+1)=
\exp(2\pi i\vec\omega\cdot\vec\tau)A_\mu(x)\exp(-2\pi i\vec\omega\cdot\vec\tau),
\ee
with $\tau_i$ the Pauli matrices.
We have chosen units such that the period in the time direction equals
one. Its proper value, where relevant, can be reinstated later on dimensional 
grounds. 

The Polyakov loop, $P(\vec x)\equiv P\exp(\int_0^1 dt A_0(x))$, is seen to 
satisfy 
\be
P_\infty\equiv\lim_{|\vec x|\rightarrow\infty}P(\vec x)=\exp(2\pi 
i\vec\omega\cdot\vec\tau).
\ee
For the periodic case ($\omega\equiv|\vec\omega|=0$) the caloron solutions are 
well known~\cite{HaSh}, but to this date no solutions for the general case 
were known.  Although it was argued that for the case of non-trivial values of 
$P_\infty$ these solutions are not important in the finite temperature 
partition function~\cite{GrPiYa}, it might be worthwhile to reinvestigate this 
issue now the solutions are known explicitly. As the finite temperature 
partition function requires the physical, i.e. gauge invariant, components of 
the fields to be periodic, we do in principle have to include also the 
configurations with $P_\infty$ non-trivial.

We will first give the explicit solution before discussing its construction.
Using a rotation, we can achieve $\hat\omega\cdot\vec\tau=\vec\omega\cdot
\vec\tau/\omega=\tau_3$, with $\omega\in[0,\half]$. The solution 
is written in terms of one real ($\phi(x)$) and one complex ($\chi(x)$)
function, and in terms of the (anti-)self dual 't~Hooft tensors~\cite{THo} 
$(\bar\eta^i_{\mu\nu})\eta^i_{\mu\nu}$ (with our conventions of $t=x_0$,
$\varepsilon_{0123}=-1$)
\be
\eta^i_{j0}=-\eta^i_{0j}=\bar\eta^i_{0j}=-\bar\eta^i_{j0}=\delta_{ij},
\quad\eta^i_{jk}=\bar\eta^i_{jk}=\varepsilon_{ijk}.
\ee
We find
\be
A_\mu(x)=\frac{i}{2}\tau_3\bar\eta^3_{\mu\nu}\partial_\nu\log\phi+\frac{i}{2}
\RE\left\{(\tau_1+i\tau_2)(\bar\eta^1_{\mu\nu}-i\bar\eta^2_{\mu\nu})
\partial_\nu\chi\right\}\phi,
\label{sol}
\ee
where
\bea
&&
\phi\!=\!\psi/\hat\psi,~\chi\!=\!\pi\rho^2\psi^{-1}e^{4\pi i\omega x_0}
\left(s^{-1}\sinh(4\pi s\omega)e^{-2\pi i x_0}+r^{-1}\sinh(4\pi r\bar\omega)
\right),~\bar\omega\!\equiv\!\half(1\!-\!2\omega),\nonumber\\&&
\hat\psi\!=\!\cosh(4\pi s\omega)\cosh(4\pi r\bar\omega)\!+\!
\frac{(r^2\!\!+\!s^2\!-\!\pi^2\rho^4)}{2rs}\sinh(4\pi s\omega)
\sinh(4\pi r\bar\omega))\!-\!\cos(2\pi x_0),\nonumber\\
&&\psi\!=\!\hat\psi+
\pi\rho^2\left(s^{-1}\sinh(4\pi s\omega)\cosh(4\pi r\bar\omega)+
r^{-1}\sinh(4\pi r\bar\omega))\cosh(4\pi s\omega)\right)
\nonumber\\ &&\qquad~~
+\frac{\pi^2\rho^4}{rs}\sinh(4\pi s\omega)\sinh(4\pi r\bar\omega).
\label{funcs}
\eea
The two radii, $r$ and $s$, that appear in eq.~(\ref{funcs}) are defined by
\be
r^2=(\vec x+2\pi\omega\rho^2\vec a)^2,\quad
s^2=(\vec x-2\pi\bar\omega\rho^2\vec a)^2,\quad \vec a=\hat\omega,
\label{radii}
\ee
and in a sense the solution can be seen as being built from a suitable 
combination of two dyons (BPS monopoles) of opposite charge, best understood 
in terms of an old construction by Taubes involving non-contractible loops in 
Yang-Mills configuration space~\cite{Taubes}. The parameter $\rho$ is related 
to the scale of the instanton solution, and the two constituent BPS 
monopoles are separated by a distance $\pi\rho^2$. Their mass ratio approaches 
$\omega/\bar\omega$ for large $\rho$, when the solution becomes static (in a 
suitable gauge). Some of these features are illustrated in figure 1. For 
$\omega=0\mod \half$, $P_\infty=\pm 1$, the gauge symmetry is no longer 
broken to the $U(1)$ subgroup generated by $\hat\omega\cdot\vec\tau$. In this 
case one of the two radii will drop out of the problem and the solution is 
spherically symmetric. The relation to the BPS monopole for large $\rho$ at 
$\omega=0$ can already be found in ref.~\cite{Rossi}. The constituent monopole 
description is also the basis for the results in ref.~\cite{Lee}, and seems to 
be the natural framework for discussing the situation for arbitrary gauge 
groups, going back to the work of Nahm~\cite{NaCal}. 

For the case $P_\infty=\pm1$ one finds $\chi=\chi^*=1-\phi^{-1}$ and 
$A_\mu=\half i\tau_j\bar\eta^j_{\mu\nu}\partial_\nu\log\phi$, which is in the 
form of the celebrated 't Hooft ansatz~\cite{THoAn}, and for which 
$\phi^{-1}\partial_\mu^2\phi=0$. For non-trivial values of $\omega$ such a 
simple characterisation is not readily available. Nevertheless, the expression
of $\tr F_{\mu\nu}^2(x)=-\partial_\mu^2\partial_\nu^2\log\phi$, derived for the 
't Hooft ansatz~\cite{THoAn}, has a remarkable generalisation to the case of 
non-trivial $\omega$, 
\be
\tr F_{\mu\nu}^2(x)=-\partial_\mu^2\partial_\nu^2\log\psi.
\label{Fsq}
\ee
This equation was used for constructing figure 1. We have also computed 
numerically the curvature directly from eq.~(\ref{sol}), checking the 
self-duality and verifying eq.~(\ref{Fsq}).

Finally we note that $\omega$ should not be considered as part of the moduli.
For each $\vec\omega$ one has a different set of solutions. This is 
particularly clear when we transform to the periodic gauge, for which 
$A_0=2\pi i\vec\omega\cdot\vec\tau$ at $|\vec x|\rightarrow\infty$. For each
choice of $\vec\omega$ we have an eight dimensional moduli space with 
as parameters the position of the caloron, which can be obtained by translating 
our solution in space and time, the scale $\rho$ and a combined rotation and 
gauge transformation (keeping $P_\infty$ fixed).

\section{The construction}

Rather than presenting the Nahm transformation~\cite{Nahm,NaCal}, we make a 
shortcut by describing the ADHM construction~\cite{ADHM}, and showing how from 
an infinite periodic array of instantons (no longer oriented parallel in group 
space~\cite{GrPiYa}) we can obtain the caloron. Input from the Nahm 
transformation comes at the point where we interpret the quaternionic valued 
matrices and vectors that appear in the ADHM construction (collectively known 
as ADHM data) as the Fourier coefficients with respect to $z$, appearing in 
the Nahm transformation. The main advantage of this approach is that already 
much is known about the calculus of multi-instantons in the ADHM 
formalism~\cite{CoGo}, in particular also for computing the metric on the 
moduli space~\cite{Osborn}. The ADHM data are obtained after applying the Nahm 
transformation. Applying this transformation for the second time yields the 
construction of the self-dual field in terms of the ADHM parameters~\cite{Buck}.

Specifically, for charge $k$ $SU(2)$ instantons the ADHM data are given
by a quaternionic valued vector $\lambda=(\lambda_1,\lambda_2,\cdots,
\lambda_k)$ and a symmetric quaternionic valued $k\times k$ matrix $B$. We 
parametrise the quaternions as linear combinations of the unit quaternions, 
$\sigma_0=1_2$ and $\sigma_j=i\tau_j$. The vector $\lambda$ is directly related
to the asymptotic behaviour of the zero-modes for the Weyl operator, which 
gives rise to the boundary terms mentioned in the introduction (see 
ref.~\cite{NaCal,CoGo,Buck} for details). This will be seen to be responsible 
for the announced delta function singularities in the case of calorons. The 
matrix $B$ is related directly to the connection for the Nahm bundle. In order 
for the ADHM data to describe a self-dual connection, they have to 
satisfy a quadratic relation which states that $B^\dagger B+\lambda^\dagger
\lambda$ is a non-singular symmetric $k\times k$ matrix whose entries are real 
(i.e. proportional to $\sigma_0$). Alternatively one may state that 
$B^\dagger B+\lambda^\dagger\lambda$ has to commute with the quaternions. 

We replace $B$ by $B-x$, with $x=x_\mu\sigma_\mu$ (a $k\times k$ unit matrix 
is implicit in our notation). The quadratic ADHM relation obviously remains 
valid.  We note that $x$ corresponds precisely to adding the flat $U(1)$ 
connection to the Nahm connection when applying the Nahm transformation for 
the {\em second} time. The self-dual gauge field is now given by~\cite{ADHM}
\be
A_\mu(x)=\frac{u^\dagger(x)(\partial_\mu u(x))-(\partial_\mu u^\dagger(x))
u(x)}{2(1+u^\dagger(x)u(x))},\quad u^\dagger(x)=\lambda(B-x)^{-1}.
\ee
There remains a redundancy which can be related to the gauge invariance 
for the Nahm bundle, 
\be
\lambda\rightarrow q\lambda T,\quad B\rightarrow T^{-1}BT,\quad A_\mu(x)
\rightarrow qA_\mu(x)\bar q,
\label{gaugetr}
\ee
where $q$ is a unit quaternion ($q\bar q=|q|^2=1$), i.e. a constant gauge 
transformation (we use $\bar x$ to denote the conjugate quaternion $x^\dagger$, 
note $\bar q=q^{-1}$), and $T$ is an orthogonal $k\times k$ matrix with real 
entries. This can be used to count the number of moduli of a charge $k$ 
instanton, being $8k-3$. Including the $q$ as moduli gives $8k$ parameters, 
forming a hyperK\"ahler manifold~\cite{DoKr,Don}.

The boundary condition $A_\mu(x+1)=\exp(2\pi i\vec\omega\cdot\vec\tau)
A_\mu(x)\exp(-2\pi i\vec\omega\cdot\vec\tau)$ is 
compatible with the algebraic nature of the ADHM construction, and can be 
implemented by
\be
\lambda_n=\exp(2\pi i n\vec\omega\cdot\vec\tau)\zeta,\quad
B_{m,n}=B_{m-1,n-1}+\delta_{m,n},
\label{per}
\ee
with $\zeta$ an arbitrary quaternion, such that
\be
u_{m+1}(x+1)=u_m(x)\exp(-2\pi i\vec\omega\cdot\vec\tau).
\label{peru}
\ee
We note that this means $k=\infty$. Indeed, $A_\mu(x)$ viewed as a solution on 
$R^4$ with unit topological charge per period has an infinite total topological
charge. For trivial holonomy ($\omega\equiv|\vec\omega|=0\mod\half$) it is 
seen that the quadratic constraint on the ADHM data is solved by choosing 
$B_{m,n}=(m+\xi)\delta_{m,n}$, with $\xi$ an arbitrary quaternion, which 
describes the position of the caloron. The caloron size is given by 
$\rho=|\zeta|$ and $\zeta/\rho$ represents a constant gauge transformation. 

The major obstacle for non-trivial holonomy was satisfying the non-linear
constraint. This can be solved most easily by introducing a Fourier 
transformation~\cite{Kraan}. Let us first give the solution in the matrix 
representation
\be
B_{m,n}=(m+\xi)\delta_{m,n}+\hat A_{m,n},\quad \hat A_{m,n}=i\bar\zeta
\hat\omega\cdot\tau\zeta\frac{\sin(2\pi\omega(m-n))}{m-n}(1-\delta_{m,n}).
\label{Ncon}
\ee
It has the right number of parameters, 8 in total, where $q\equiv\zeta/\rho$ 
is split in a $U(1)$ part commuting with $P_\infty$, describing the residual 
$U(1)$ gauge invariance, and a part $SU(2)/U(1)$ describing a rotation of the 
vector $\vec\omega$, compensated by a gauge transformation to ensure that 
$P_\infty$, or equivalently the periodicity condition, is unaltered.

It is advantageous to first give some general results, valid for arbitrary 
instantons, which seem to be new and give a more efficient way of representing 
$A_\mu(x)$. The derivation is straightforward and will presented elsewhere. 
It is well known that in this problem two Green's functions 
appear~\cite{NaCal,CoGo}. One is associated to the quadratic ADHM 
relation
\be
f_x=(\Delta^\dagger(x)\Delta(x))^{-1},\quad\Delta^\dagger(x)=
(\lambda^\dagger,(B-x)^\dagger),\quad
\Delta^\dagger(x)\Delta(x)=(B-x)^\dagger(B-x)+\lambda^\dagger\lambda.
\ee
$\Delta^\dagger(x)$ is a $k\times(k+1)$ dimensional quaternionic matrix.
Self-duality implies that $f_x$ commutes with the quaternions. 
The other Green's function is given by
\be
G_x=((B-x)^\dagger(B-x))^{-1}.
\ee
From the definition of $u(x)$ it follows that 
\be
\phi\equiv 1+\lambda G_x\lambda^\dagger=1+u^\dagger(x)u(x).
\ee
One finds the following compact result
\be 
A_\mu(x)=-\half\phi\partial_\nu\left(\phi^{-1}(\lambda\bar\eta_{\nu\mu}G_x
\lambda^\dagger)\right),
\ee
where $\bar\eta_{\mu\nu}\equiv\sigma_i\bar\eta^i_{\mu\nu}$ (the 't Hooft 
tensors~\cite{THo} may be defined through $\bar\eta_{\mu\nu}=\half(\bar
\sigma_\mu\sigma_\nu-\bar\sigma_\nu\sigma_\mu)$ and $\eta_{\mu\nu}=
\half(\sigma_\mu\bar\sigma_\nu-\sigma_\nu\bar\sigma_\mu)$).

By choosing $B$ diagonal and the entries of $B$ and $\lambda$ real, this result 
immediately leads to the subclass of solutions that are expressed in terms 
of the 't Hooft ansatz~\cite{THoAn}. The quadratic ADHM condition is obviously 
satisfied, and $A_\mu(x)=\half \bar\eta_{\mu\nu}\partial_\nu\log\phi$. 

We note that the Green's functions $f_x$ and $G_x$ are intimately related. 
In particular
\be
G_x\lambda^\dagger=\phi f_x\lambda^\dagger,
\ee
which implies that 
\be
\phi=(1-\lambda f_x\lambda^\dagger)^{-1},\quad
A_\mu(x)=-\half\phi\partial_\nu\left(\lambda\bar\eta_{\nu\mu}f_x
\lambda^\dagger\right).
\label{Asim}
\ee
As a consequence we will only need to know $f_x$. This Green's function is 
simpler to determine than $G_x$, since $f_x$ is a proportional to $\sigma_0$. 
A standard computation, that lies at the heart of showing that the curvature 
obtained after applying the Nahm transformation is 
self-dual~\cite{NaCal,CoGo,BrBa} (for which it is crucial that $f_x$ commutes 
with the quaternions), yields
\be
F_{\mu\nu}=2\phi^{-1}u^\dagger\eta_{\mu\nu}f_xu.
\label{curv}
\ee

We have now reduced the explicit computation of instanton solutions to the 
computation of $f_x$. Incidentally, it can be verified that all infinite sums 
involved in expressions that appear for the calorons are convergent. 
Eq.~(\ref{curv}) demonstrates that it gives a self-dual solution. We now 
discuss the Fourier transformation, in terms of which one solves for the 
quadratic ADHM constraint and for the Green's function $f_x$. One defines
\bea
&&\hat\lambda(z)=\sum_m\exp(2\pi imz)\lambda_m=(P_+\delta(z-\omega)+P_-\delta(
z+\omega))\zeta, \nonumber\\
&&\delta(z^\prime-z)\hat D(z)=\sum_{m,n}\exp\left(2\pi i(mz^\prime-nz)\right)
B_{m,n},
\eea
where $P_\pm=\half(1\pm\hat\omega\cdot\vec\tau)$. Parametrising $B_{m,n}$ as 
before in terms of $\xi$ and $\hat A_{m,n}$, with $\delta(z^\prime-z)\hat A(z)
=\sum_{m,n}\exp(2\pi i(mz^\prime-nz))\hat A_{m,n}$, we find
\be
\hat D(z)=\frac{1}{2\pi i}\frac{d}{dz}+\xi+\hat A(z).
\ee
Thus, $B$ has been turned into a differential operator, precisely the Weyl 
operator appearing in the Nahm transformation, with $\hat A(z)\equiv\hat 
A_\mu(z)\sigma_\mu$ the connection for the Nahm bundle~\cite{Buck} (up to 
factors $2\pi i$, to match with the conventions of the ADHM construction). The 
Nahm transformation would require $\hat D^\dagger\hat D$ to commute with the
quaternions, which is equivalent to saying that the curvature of the Nahm
connection is self-dual. Due to the boundary terms discussed in the 
introduction, this self-duality is violated at a finite number of 
points~\cite{Buck} and the presence of $\lambda^\dagger\lambda$ in the 
quadratic ADHM relation is precisely so as to correct for the violations 
of self-duality, in accordance with the expectations expressed in the 
introduction. After Fourier transformation this quadratic relation reads, 
with a slight abuse of notation,
\be
(\Delta^\dagger(x)\Delta(x))(z)\equiv(\hat D(z)-x)^\dagger(\hat D(z)-x)
+\hat\Lambda(z),
\ee
where
\be
\delta(z^\prime-z)\hat\Lambda(z)=\hat\lambda^\dagger(z^\prime)\hat\lambda(z),
\quad\hat\Lambda(z)=\bar\zeta\left(P_+\delta(z-\omega)+P_-\delta(z+\omega)
\right)\zeta.
\ee
The condition that $\Delta^\dagger(x)\Delta(x)$ has to commute with the 
quaternions is now seen to lead to the equation
\be
d\hat A(z)/dz=\pi\bar\zeta\hat\omega\cdot\vec\sigma\zeta(\delta(z+\omega)-
\delta(z-\omega)),
\ee
which is solved by (eliminating an arbitrary additive constant that can be
absorbed in $\xi$ by imposing $\int^1_0 dz\hat A(z)=0$),
\be
\hat A(z)=\bar\zeta\hat\omega\cdot\vec\sigma\zeta\hat A^{(0)}(z),\quad
\hat A^{(0)}(z)=\pi(1-2\omega-\chi_\omega(z)),
\ee
where $\chi_\omega(z)=1$ for $\omega<z<1-\omega$ (requiring 
$\omega\in[0,\half]$) and 0 elsewhere. Fourier transformation of $\hat A(z)$ 
yields the result in eq.~(\ref{Ncon}). 

As $x$ and $\xi$ always occur in the combination $x-\xi$, we absorb $\xi$ by a 
translation in $x$. The computation of $f_x$ now reduces to a one-dimensional 
quantum mechanical problem on the circle,
\be
\left\{\left(\frac{1}{2\pi i}\frac{d}{dz}-x_0\right)^2\!+\!r^2\chi_\omega(z)\!
+\!s^2(1\!-\!\chi_\omega(z))\!+\!\half\rho^2(\delta(z\!+\!\omega)\!+
\!\delta(z\!-\!\omega))\right\}f_x(z,z^\prime)=\delta(z-z^\prime),
\ee
where the radii $r$ and $s$ were given in eq.~(\ref{radii}) (note that here 
$\vec a\!\cdot\!\vec\sigma=\bar\zeta\hat\omega\!\cdot\!\vec\sigma\zeta$).
We will present the explicit analytic solution for $f_x(z,z^\prime)$ elsewhere,
but it should be noted that, due to the particular form of $\hat\lambda(z)$,
only $f_x(\omega,\omega)=f_x(-\omega,-\omega)$ and $f_x(\omega,
-\omega)=f_x(-\omega,\omega)^*$, respectively real and complex functions
of $x_\mu$, will occur in the evaluation of $A_\mu(x)$ (see eq.~(\ref{Asim})).
To obtain eq.~(\ref{sol}) from eq.~(\ref{Asim}), one moves $\bar\eta_{\mu\nu}$ 
through $\hat\lambda(z)$, which for non-trivial $\omega$ do not commute.

We close this section by quoting a useful and remarkable 
result~\cite{CoGo,Osborn},
\be
\tr F_{\mu\nu}^2(x)=-\partial_\mu^2\partial_\nu^2\log\det(f_x),
\ee
which leads to the result given in eq.~(\ref{Fsq}). Although for the caloron 
$\log\det(f_x)$ is divergent, $\partial_\mu\log\det(f_x)$ is well defined.

\section{The geometry of moduli space}

The moduli space of the self-dual solutions is given by the ADHM data, or 
equivalently by the Nahm connections. The latter are self-dual connections on 
the dual space and the Nahm transformation provides precisely a 
T-duality~\cite{Tdual}. It has been well established that this transformation
preserves the metric and hyperK\"ahler structure of the moduli 
spaces~\cite{BrBa,DoKr}. By computing this metric on the moduli space we can 
determine its geometry. We use results due to Osborn~\cite{Osborn}, which can 
be readily transposed to the case of the calorons and become particularly 
elegant after the Fourier transformation.

Since we have a closed expression for $A_\mu(x)$ in terms of the ADHM 
parameters, we can compute the variations $\delta A_\mu(x)$ with respect to 
the moduli in terms of variations of the ADHM data, summarised in terms of 
$\delta\Delta(x)$.  The metric is obtained by computing $||\delta A||^2=
-\int d_4 x\tr(P\delta A_\mu(x))^2$, where $P$ is the projection on the 
transverse gauge fields, achieved by applying an infinitesimal gauge 
transformation such that $\delta A^\prime_\mu(x)$ satisfies the background 
gauge condition $D_\mu\delta A^\prime_\mu(x)=0$. It can be shown 
that~\cite{Osborn}
\be
D_\mu\delta A_\mu(x)=\phi^{-1}u^\dagger f_x\sigma_\mu\left((\delta
\Delta^\dagger)\Delta-\Delta^\dagger(\delta\Delta)\right)\bar\sigma_\mu f_xu,
\ee
which vanishes if and only if $\half\tr\left((\delta\Delta^\dagger)\Delta-
\Delta^\dagger(\delta\Delta)\right)=0$. This condition is precisely the
background gauge condition for the Nahm connection (for $T^4$ this is an 
exact statement~\cite{BrBa}, whereas in general $\lambda$ provides the 
corrections due the asymptotic behaviour of the chiral zero-modes of the Weyl
operator). In particular this implies that the background gauge condition is 
preserved under the Nahm, or T-duality, transformation. Projection to a 
transverse variation of the connection can therefore be achieved by applying 
an infinitesimal gauge transformation to the ADHM data as given in 
eq.~(\ref{gaugetr}) ($q=1$). Under such an infinitesimal gauge transformation, 
$T=\exp(\delta X)=1+\delta X+\cdots$, 
\be
\delta_X\lambda=\lambda \delta X,\quad \delta_X B=[B,\delta X],\quad 
\delta X^\dagger=-\delta X.
\ee
Replacing $\delta\Delta$ by $C_X\equiv\delta\Delta+\delta_X\Delta$, the
background gauge condition gives an equation for $\delta X$ in terms of 
$\delta\Delta$,
\be
\half\tr\left(B^\dagger[B,\delta X]-[B^\dagger,\delta X]B+2\delta X\Lambda
+\Delta^\dagger\delta\Delta-\delta\Delta^\dagger\Delta\right)=0.
\ee
For the caloron, with $\delta X$ preserving the periodicity (\ref{peru}), the 
transformation of $\delta\Delta$ to a transverse variation can now be 
reformulated after Fourier transformation as
\be
-\frac{1}{4\pi^2}\frac{d^2\delta \hat X(z)}{dz^2}+|\zeta|^2(\delta(z-\omega)+
\delta(z+\omega))\delta\hat X(z)=\frac{i}{4}\tr\left((\delta\zeta
\bar\zeta-\zeta\delta\bar\zeta)\hat\omega\cdot\vec\sigma\right)
(\delta(z-\omega)-\delta(z+\omega)),
\ee
where $\delta(z^\prime-z)\delta\hat X(z)=\sum_{m,n}\exp(2\pi i(mz^\prime-nz))
\delta X_{m,n}$. Solving for $\delta\hat X(z)$ gives
\be
\delta\hat X(z)=-\pi i\frac{\tr\left((\delta\zeta\bar\zeta-\zeta\delta\bar\zeta)
\hat\omega\cdot\vec\sigma\right)}{1+4\pi^2\omega(1-2\omega)|\zeta|^2}
\int_0^zdz^\prime\hat A^{(0)}(z^\prime),
\ee
which is a zig-zag function (periodic and odd in $z$), with discontinuous 
derivatives at $z=\pm\omega$.

The following miraculous formula due to Osborn~\cite{Osborn} allows us to 
compute $||\delta A||^2$,
\be
\tr(P\delta A_\mu(x))^2=\half\partial_\mu^2\tr\Tr\left(C_X^\dagger(2-\Delta(x)
f_x\Delta^\dagger(x))C_Xf_x\right).
\ee
Since the right-hand side of this equation is smooth and a total derivative,
the integration over space and time is completely determined by the behaviour
for $r=|\vec x|\rightarrow\infty$. In this limit we may replace $f_x(z,
z^\prime)$ by $\pi r^{-1}\exp\left(-2\pi r|z-z^\prime| +2\pi ix_0(z-z^\prime)
\right)$ (near $z=z^\prime$, properly extended as a function on $S^1\times 
S^1$). From this we find
\be
||\delta A||^2=2\pi^2\tr\int^1_0dz\left\{d\hat B^\dagger(z)d\hat B(z)+
2d\hat\lambda^\dagger(z)\int_0^1dz^\prime d\hat\lambda(z^\prime)\right\},
\ee
where
\bea
d\hat\lambda(z)&\equiv&P_+\delta(z-\omega)(\delta\zeta+\zeta \delta
\hat X(\omega))+P_-\delta(z+\omega)(\delta\zeta-\zeta \delta\hat X(\omega)),
\nonumber\\ d\hat B(z)&\equiv&\delta\xi+\delta\hat A(z)+\frac{1}{2\pi i}
\frac{d\delta \hat X(z)}{dz}.
\eea
For the metric one finds the explicit result
\be
||\delta A||^2=4\pi^2|\delta\xi|^2+8\pi^2(1+R^2)|\delta\zeta|^2-2\pi^2R^2
|\zeta|^2(1+\frac{1}{1+R^2})(\hat\omega\cdot\delta\vec\sigma)^2,
\ee
where ($\zeta\equiv y_\mu\sigma_\mu$)
\be
R^2=\pi^2|\zeta|^2/M^2,\quad M^{-2}=8\omega(1-2\omega),\quad\half\delta\sigma_j
=\eta^j_{\mu\nu}|\zeta|^{-2}y_\mu dy_\nu.
\ee
One readily recognises, putting $\delta\xi=0$, the Taub-NUT metric~\cite{Nut} 
with mass $M$. For $\hat\omega$ in the third direction this metric is given 
by~\cite{Haw}
\be
ds^2=(1+\frac{x^2}{16M^2})\left(dx^2+\quart x^2(d\sigma_1^2+d\sigma_2^2)\right)+
\quart x^2 d\sigma_3^2/(1+\frac{x^2}{16M^2}),
\ee
where we identify $x^2=8\pi^2\rho^2$. We note that the Taub-NUT space is a 
self-dual Einstein manifold~\cite{GiPo} and that it has a hyperK\"ahler 
structure~\cite{AtHi}, inherited from the hyperK\"ahler structure 
of $R^3\times S^1$.

\section{Conclusions}

We have found the explicit charge one $SU(2)$ caloron solutions with the 
Polyakov loop at spatial infinity non-trivial. Previously only solutions for 
which the latter was trivial were known~\cite{HaSh}. Those were argued to 
dominate in the instanton contribution to the finite temperature partition 
function~\cite{GrPiYa}, a question that can now more directly be addressed 
and is perhaps of physical significance.

We have shown that the moduli space of these solutions forms a Taub-NUT space, 
providing an exact classical T-duality between H-monopoles and Kaluza-Klein 
monopoles~\cite{GrHaMo}. Indeed it is well-known that the Taub-NUT metric 
describes (the spatial part of) the Kaluza-Klein monopole~\cite{SoGrPe} with 
compactification radius $4M$. Most importantly we have related the holonomy to 
the compactification radii involved in the dual descriptions.

\section*{Acknowledgements}

This work was started when one of us (PvB) was visiting the Newton Institute
during the first half of 1997. He thanks the staff for their hospitality 
and gratefully acknowledges discussions with Jerome Gauntlett, Nick Manton, 
Werner Nahm, Hiraku Nakajima, David Olive and Erik Verlinde. TCK was supported 
by a grant from the FOM/SWON Association for Mathematical Physics.

\newpage
\begin{figure}[htb]
\vspace{18cm}
\includegraphics{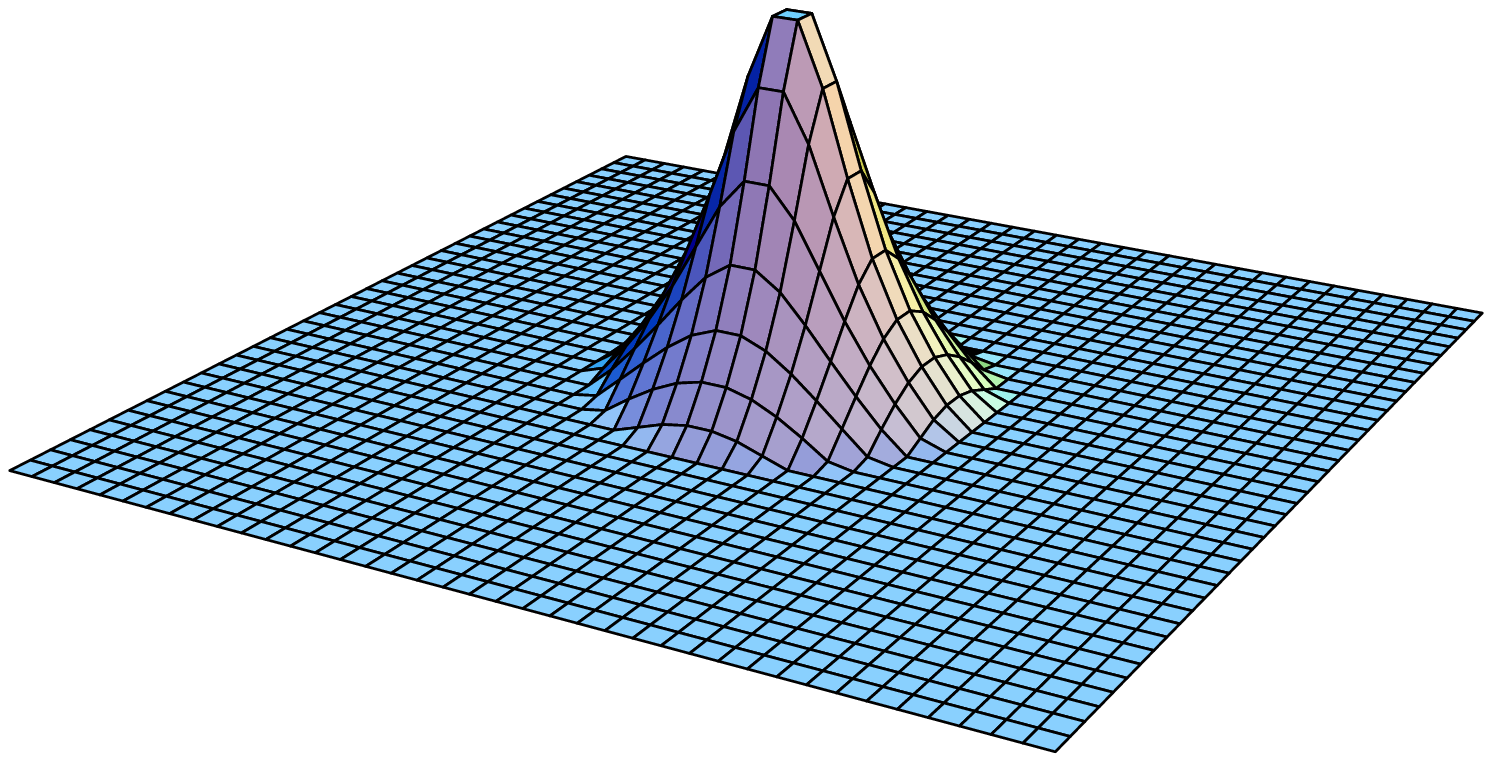}
\includegraphics{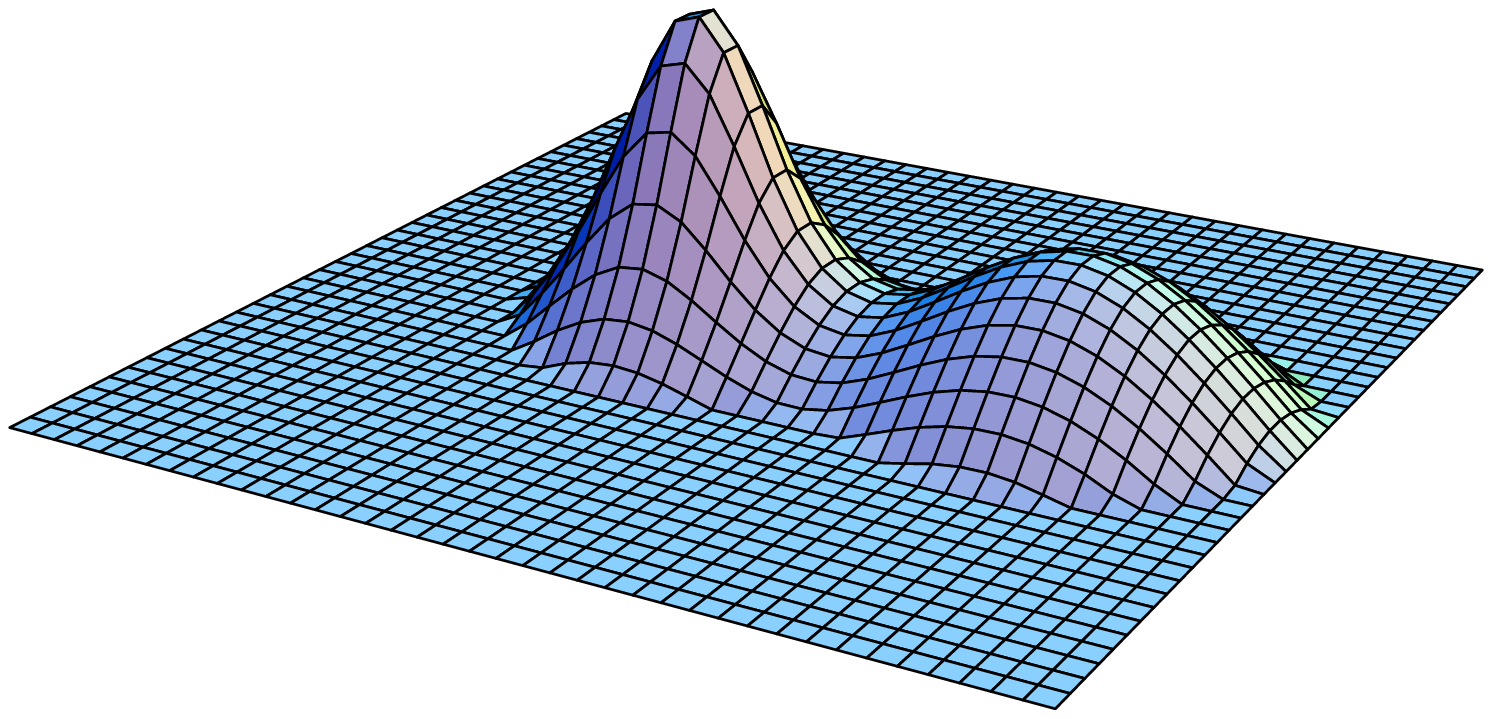}
\includegraphics{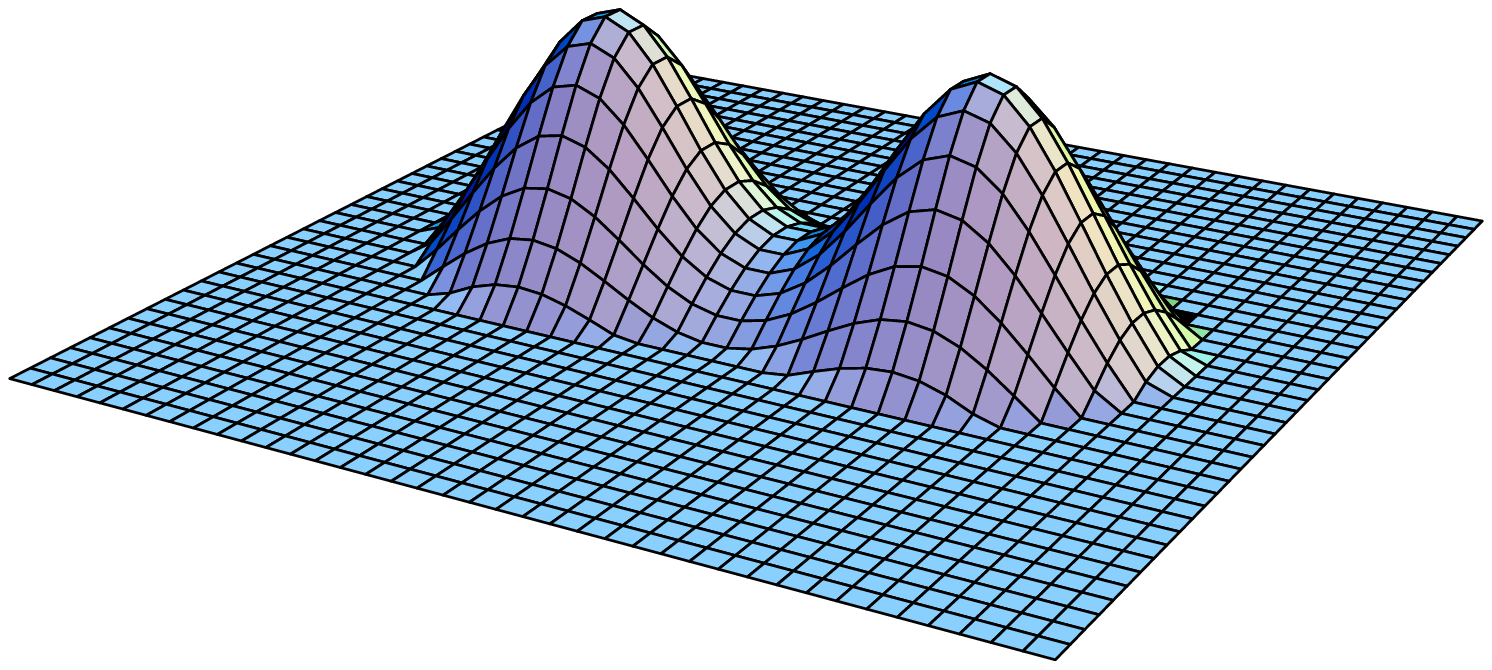}
\caption{Profiles for calorons at $\omega=0$, 0.125, 0.25 (from top
to bottom) with $\rho=1$. The axis connecting the lumps, separated by a 
distance $\pi$ (for $\omega\neq0$), corresponds to the direction of 
$\hat\omega$. The other direction indicates the distance to this axis, 
making use of the axial symmetry of the solutions. Vertically is 
plotted the action density, at the time of its maximal value, on equal 
logarithmic scales for the three profiles. The profiles were cut off at an 
action density below $1/e$. The mass ratio of the two lumps is approximately 
$\omega/\bar\omega$, i.e. zero (no second lump), a third and 
one (equal masses), for the respective values of $\omega$.}
\end{figure}
\end{document}